\documentclass[12pt]{article}

\usepackage{amssymb}
\usepackage{amsmath}
\usepackage{amscd}
\usepackage{latexsym}
\usepackage{graphicx}

\usepackage{cite}

\newcommand{\be}{\begin{equation}}
\newcommand{\ee}{\end{equation}}

\newcommand{\dlt}{\delta}

\newcommand{\bt}{\beta}
\newcommand{\vp}{\varphi}
\newcommand{\ep}{\varepsilon}
\newcommand{\al}{\alpha}
\newcommand{\ra}{\rightarrow}

\newcommand{\dgr}{\dagger}
\newcommand{\lbd}{\lambda}

\newcommand{\rgl}{\rangle}
\newcommand{\lgl}{\langle}
\newcommand{\cH}{{\cal H}}
\newcommand{\cL}{{\cal L}}
\newcommand{\cD}{{\cal D}}

\begin{document}

\begin{center}
 
{\Large{\bf Measure of entanglement production by quantum operations} \\ [5mm]

V.I. Yukalov$^{1,2}$ and E.P. Yukalova$^{3}$}  \\ [3mm]

{\it
$^1$Bogolubov Laboratory of Theoretical Physics, \\
Joint Institute for Nuclear Research, Dubna 141980, Russia \\ [2mm]

$^2$Instituto de Fisica de S\~ao Carlos, Universidade de S\~ao Paulo, \\
CP 369, S\~ao Carlos 13560-970, S\~ao Paulo, Brazil \\ [2mm]

$^3$Laboratory of Information Technologies, \\
Joint Institute for Nuclear Research, Dubna 141980, Russia } \\ [3mm]

{\bf E-mails}: {\it yukalov@theor.jinr.ru}, ~~ {\it yukalova@theor.jinr.ru} \\

\end{center}

\vskip 1cm

\begin{abstract}

A measure of entanglement production by quantum operations is suggested. This measure 
is general, being valid for operations over pure states as well as over mixed states, 
for equilibrium as well as for nonequilibrium processes. The measure of entanglement 
production satisfies all properties typical of such a characteristic. Systems of 
arbitrary nature can be treated, described by field operators, spin operators, or any 
other operators, which is realized by defining generalized correlation matrices. 
Particular cases of entanglement production are considered.
         
\end{abstract}

\vskip 2mm

\section{Entangled structures}

The notion of entanglement appears as an important ingredient in quantum information 
processing 
\cite{Williams_1,Nielsen_2,Vedral_3,Keyl_4,Horodecki_5,Guhne_6,Yukalov_7,Wilde_8}. 
This notion was introduced by Schr\"{o}dinger \cite{Schrodinger_9,Schrodinger_10} for
quantum systems that can be separated into several subsystems. One has to distinguish 
two different types of entanglement, entangled structures and entangling operations.  
In order to better understand the following, let us, first, recall the definition of
entangled structures, such as entangled functions and entangled states (statistical 
operators).  

Suppose one considers a composite system consisting of several subsystems enumerated
by the index $i = 1,2,\ldots,N$. Each subsystem is characterized by a Hilbert space
\be
\label{1}
   \cH_i \; = \; \overline\cL\{ \; |\; \vp_i \; \rgl \; \}
\ee
that is a closed linear envelope over an orthonormal basis. The total system space is
the tensor product
\be
\label{2}
\cH \; = \; \bigotimes_{i=1}^N \cH_i \; = 
\; \overline \cL \{ \; \bigotimes_{i=1}^N |\; \vp_i \; \rgl \; \} \;   .
\ee
Among the members of the total space there are functions of two types, separable 
functions of the product form
$$
|\; \vp_{sep}^\al \; \rgl  \; = \;   \bigotimes_{i=1}^N |\; \vp_{i\al} \; \rgl
\qquad
(\; |\; \vp_{i\al} \; \rgl \in \cH_i \; )
$$
and entangled functions 
$$
|\; \vp_{ent} \; \rgl  \; = \;   \sum_\al c_\al \; |\; \vp_{sep}^\al \; \rgl
\qquad
(\; |\; \vp_{sep}^\al \; \rgl \in \cH \; ) \;  ,
$$
where at least one $c_\alpha \neq 0$.  

Respectively, the total Hilbert space (\ref{2}) consists of two types of sets, 
a separable, or disentangled, set and an entangled set,
\be
\label{3}
\cD \; = \; \{\; |\; \vp_{sep} \; \rgl \; \} \; , \qquad
\cH \setminus \cD \; = \; \{ \; |\; \vp_{ent} \; \rgl \; \} \;  ,
\ee
so that
\be
\label{4}
\cH \; = \; \cD \;  \bigcup \; \cH \setminus \cD \;   .
\ee

Statistical operators, or states, also can be separable, or entangled. The separable 
states can be pure, of the form
\be
\label{5}
\hat\rho_{sep}^\al \; = \; |\; \vp_{sep}^\al \; \rgl \lgl \; \vp_{sep}^\al \; | \; = \;
\bigotimes_{i=1}^N  |\; \vp_{i\al} \; \rgl \lgl \; \vp_{i\al} \; |  ,
\ee
and mixed, of the form
\be
\label{6}
\hat\rho_{sep} \; = \; \sum_\al \lbd_\al \; \hat\rho_{sep}^\al \; = \; 
\sum_\al \lbd_\al \bigotimes_{i=1}^N \hat\rho_{i\al} \;  .
\ee
The states that cannot be represented in these forms, are named entangled.

\section{Entangling operations} 

The operations $\hat{A}$, acting on the Hilbert space (\ref{2}), can be of two types,
nonentangling, when they do not induce entangled states from the separable states, 
\be
\label{7}
 \hat A \cD \; = \; \cD \qquad (nonentangling) \;  ,   
\ee
and entangling, when, acting on separable states, they produce entangled states 
\be
\label{8}
 \hat A \cD \; = \; \cH \; \setminus \; \cD \qquad (entangling) \;    .
\ee
The question of principal interest is: How to measure the entangling power of quantum 
operations?

There have been attempts of defining an entanglement probability by taking the matrix
element squared, for the considered operator and fixed separable and entangled functions 
from the given Hilbert space, proportional to 
$|\langle \varphi_{ent}|\hat{A}|\varphi_{sep} \rangle|^2$. However, as is evident,
this form does not represent a general entangling power over the whole Hilbert space.

The general measure of entanglement production for a quantum operation ${\hat A}$ was 
advanced in \cite{Yukalov_11,Yukalov_12,Yukalov_13}. Here we recall the main definition
and properties of this measure and suggest a generalization of the notion for measuring
entangling quantum correlations of generalized correlation functions. First, we need the
following result.

\vskip 2mm

{\bf Theorem}. For each trace-class operator $\hat{A}$ acting on a Hilbert space
$\mathcal{H}$ there exists a unique separable operator $\hat{A}^\otimes$ such that it 
is nonentangling,
\be
\label{9}
  \hat A^\otimes \cD \; = \; \cD \qquad (nonentangling) \;   ,
\ee
and trace-invariant,
\be
\label{10}
 {\rm Tr}_\cH \hat A^\otimes \; = \; {\rm Tr}_\cH \hat A \qquad (trace-invariant) \;    ,
\ee
having the form
\be
\label{11}
 \hat A^\otimes \; = \; 
\frac{\bigotimes_{i=1}^N \hat A_i}{({\rm Tr}_\cH \hat A)^{N-1} } \; ,  
\ee
in which
\be
\label{12}
 \hat A_i \; \equiv \; {\rm Tr}_{\cH\setminus\cH_i} \; \hat A \;  .
\ee

\vskip 2mm

We also need to define the norm of the operator $\|\hat{A}\|$. This can be, e.g., the 
Hilbert-Schmidt (Frobenius) norm or the operator norm
$$
|| \; \hat A \; || \; = \; 
\sup_\vp \frac{ || \; \hat A \vp\; ||}{|| \; \vp \; || } 
\qquad
( \vp \equiv |\; \vp \; \rgl \in \cD ) \;  .
$$ 
In the latter, the nonzero functions are from the disentangled set $\mathcal{D}$. For 
Hermitian operators, the operator norm reduces to the Hermitian norm
\be
\label{13}
 || \; \hat A \; || \; = \; 
\sup_\vp \frac{ \lgl \; \vp \; | \; \hat A \; | \; \vp \; \rgl}{|| \; \vp \; || } 
\qquad
( \hat A^+  = \hat A ) \;  .
\ee
       
The measure of entanglement produced by the operator $\hat{A}$ is defined as
\be
\label{14}
\ep(\hat A) \; = \; 
\log \; \frac{|| \; \hat A \; ||}{|| \; \hat A^\otimes \; || } \;  ,
\ee
with the logarithm over any base that is convenient. This measure quantifies the amount
of entanglement produced by a quantum operation under the action on a set of disentangled
functions. The operator produces entanglement in the functions due to its structure
incorporating quantum correlations.

The measure of entanglement production enjoys all properties required for representing 
a measure, being: \\
(i) semi-positive,
\be
\label{15}
 \ep(\hat A) \; \geq \; 0 \;    ,
\ee
(ii) zero for separable operators,
\be
\label{16}
 \ep(\hat A^\otimes) \; = \; 0 \;  ,
\ee
(iii) continuous by norm with respect to parameters,
\be
\label{17}
 \ep(\hat A_t) \; \ra \;  \ep(\hat A_0)
\qquad 
( ||\hat A_t|| \ra ||\hat A_0|| , ~ t \ra 0 ) \;  ,
\ee
(iv) additive,
\be
\label{18}
 \ep(\hat A \bigotimes \hat B) \; = \; \ep(\hat A) + \ep(\hat B) \;  ,
\ee
(v) invariant under local unitary operations,
\be
\label{19}
\ep(\hat U^+ \hat A \; \hat U ) \; = \; \ep(\hat A)  \qquad
\left( \hat U = \bigotimes_{i=1}^N \hat U_i \right) \; .
\ee

\section{Pure states}

To illustrate the use of the defined measure, let us start with simple quantum systems 
\cite{Adams_14}, which are described by pure quantum states. For instance, the 
Einstein-Podolsky-Rosen state
\be
\label{20}
\hat \rho_{EPR} \; = \; |\; \vp_{EPR} \; \rgl \lgl \; \vp_{EPR}\; |
\ee
is formed by the functions
\be
\label{21}
 |\; \vp_{EPR} \; \rgl \; = \; 
c_{12} \; | \; 12 \; \rgl + c_{21} \; | \; 21 \; \rgl \; ,
\ee
in which
$$
|\; c_{12}\; |^2 + |\; c_{21} \; |^2 \; = \; 1 \; , 
\qquad
|\; n \; \rgl \; \equiv \; |\; \vp_n \; \rgl \;   .
$$
The related separable state is
\be
\label{22}
 \hat \rho_{EPR}^\otimes \; = \; \hat\rho_1 \bigotimes \hat\rho_2 \; ,
\ee
where
$$
 \hat\rho_1 \; = \; |\; c_{12}\; |^2  \; | \; 1 \; \rgl \lgl \; 1 \; | +
|\; c_{21}\; |^2  \; | \; 2 \; \rgl \lgl \; 2 \; | \; = \; \hat\rho_2 \; .
$$
The entanglement production measure reads as
\be
\label{23}
 \ep(\hat\rho_{EPR} ) \; = \; 
- \log \; \sup \; \{ |\; c_{12}\; |^2  , ~  |\; c_{21}\; |^2 \}
\ee
reaching the maximum at 
\be
\label{24}
\ep(\hat\rho_{EPR} ) \; = \; \log 2
 \qquad  
\left( |\; c_{12}\; |^2 = \frac{1}{2}\right) \;  .
\ee

Similarly, for the Bell state
\be
\label{25}
 \hat\rho_B  \; = \; | \; \vp_B \; \rgl \lgl \; \vp_B \; | \; ,
\ee
with
\be
\label{26}
 | \; \vp_B \; \rgl  \; = \; 
c_{11} \; | \; 11 \; \rgl + c_{22} \; | \; 22 \; \rgl \;  ,
\ee
one has
$$
 \hat\rho_B^\otimes \; = \;  \hat\rho_1 \bigotimes \hat\rho_2 \; ,
\qquad
\hat\rho_1 \; = \; |\; c_{11}\; |^2 \; |\; 1 \; \rgl \lgl \; 1 \; | +
|\; c_{22}\; |^2 \; |\; 2 \; \rgl \lgl \; 2 \; | \; .
$$

The entanglement production measure is
\be
\label{27}
 \ep(\hat\rho_B ) \; = \; 
- \log \; \sup \; \{ |\; c_{11}\; |^2  , ~  |\; c_{22}\; |^2 \} \; ,
\ee
with its maximum at 
\be
\label{28}
 \ep(\hat\rho_B ) \; = \; \log 2
 \qquad  
\left( |\; c_{11}\; |^2 = \frac{1}{2}\right) \;  .
\ee

For the multicat state
\be
\label{29}
\hat\rho_{MC}  \; = \; | \; \vp_{MC} \; \rgl \lgl \; \vp_{MC} \; | \;   ,
\qquad
 | \; \vp_{MC} \; \rgl  \; = \; 
c_1 \; | \; 11\ldots 1 \; \rgl + c_2 \; | \; 22\ldots 2 \; \rgl \;  ,
\ee
the measure is
\be
\label{30}
\ep(\hat\rho_{MC} ) \; = \; (1 - N) \;
\log \; \sup \; \{ |\; c_{1}\; |^2  , ~  |\; c_{2}\; |^2 \} \; ,
\ee
having the maximum at
\be
\label{31}
\ep(\hat\rho_{MC} ) \; = \; ( N - 1) \log 2
 \qquad  
\left( |\; c_{1}\; |^2 = \frac{1}{2}\right) \;   .
\ee

For the multimode state, 
\be
\label{32}
\hat\rho_{MM}  \; = \; | \; \vp_{MM} \; \rgl \lgl \; \vp_{MM} \; | \;   ,
\qquad
 | \; \vp_{MM} \; \rgl  \; = \; \sum_n 
c_n \; | \; n n \ldots n \; \rgl \;   ,
\ee
one has the measure
\be
\label{33}
 \ep(\hat\rho_{MM} ) \; = \; (1 - N) \;
\log \; \sup_n \; \{\; |\; c_{n}\; |^2 \; \} \;   ,
\ee
whose maximum is
\be
\label{34}
\ep(\hat\rho_{MM} ) \; = \; ( N - 1) \log M
 \qquad  
\left( |\; c_{n}\; |^2 = \frac{1}{M}\right) \;    .
\ee

The Hartree-Fock state
\be
\label{35}
\hat\rho_{HF}  \; = \; | \; \vp_{HF} \; \rgl \lgl \; \vp_{HF} \; | \;   ,
\qquad
 | \; \vp_{FH} \; \rgl  \; = \; \frac{1}{\sqrt{N!} } \sum_{sym}
 | \; 1 2 \ldots N \; \rgl 
\ee
gives the measure
\be
\label{36}
  \ep(\hat\rho_{HF} ) \; = \; \log \; \frac{N^N}{N!} \; .
\ee
At large $N$, we get
\be
\label{37}
\ep(\hat\rho_{HF} ) \; \simeq \; N \log e \qquad ( N \gg 1) \;  .
\ee
And for two particles, we have
$$
\ep(\hat\rho_{HF} ) \; = \;  \log 2 \qquad ( N = 2) \;   .
$$

\section{Mixed states}

As an example of mixed states, let us consider thermal entanglement produced by means 
of a statistical operator
\be
\label{38}
 \hat\rho \; = \; \frac{1}{Z} \; e^{-\bt H} 
\qquad 
\left( Z = {\rm Tr}\; e^{-\bt H} \right) \;  .
\ee
For a binary Ising register, with the Hamiltonian
\be
\label{39}
H \; = \; - \frac{1}{2} \sum_{i\neq j} J_{ij} S_i^z S_j^z - B \sum_i S_i^z \;  ,
\ee
in which $i,j = 1,2$, the measure can be calculated explicitly 
\cite{Yukalov_12,Yukalov_15,Yukalov_16}. In the limiting cases, we find 
$$
\lim_{B\ra 0}\ep(\hat\rho) \; = \; 
\log \; \frac{e^{|\bt J S^2|}}{\cosh(\bt J S^2)}  \; ,
$$
$$
\lim_{T\ra 0} \ep(\hat\rho) \; = \; \log 2 \; ,
$$
\be
\label{40}
 \lim_{B\ra \infty} \ep(\hat\rho) \; = \; 
\lim_{T\ra \infty} \ep(\hat\rho) \; = \; 0 \; .
\ee
Notice that quantum correlations in the register are due to interactions, without which
there is no entanglement production,
\be
\label{41}
 \lim_{J_{ij}\ra 0} \ep(\hat\rho) \; = \;  \ep(\hat\rho^\otimes) \; = \; 0 \;  .
\ee  

The evolution operator $\hat{U}^{-iHt}$ also can produce entanglement 
\cite{Yukalov_17,Yukalov_18}. The related entanglement production is given by the 
measure
\be
\label{42}
\ep(\hat U(t) ) \; = \; \log\; \frac{||\; \hat U(t)\;||}{||\; \hat U^\otimes(t)\; ||} \;   .
\ee

\section{Correlation entanglement}

The suggested measure quantifies the amount of entanglement produced in a quantum
system by a quantum operation. The entanglement production is due to the structure
of the related quantum operator that, acting on separable functions, produces entangled
functions. This entanglement is realized because the considered operator contains 
quantum correlated elements inducing correlations into the functions, which the operator 
acts on. In the previous sections, simple quantum systems are treated in order to 
clearly illustrate the process of inducing entanglement. The process can be generalized
to complicated quantum statistical systems described not by so simple states, for which 
the introduced measure also could quantify the amount of entanglement and quantum
correlations.   

The usual understanding of entanglement is applied to quantum states, that is, to 
quantum statistical operators, while the measure (\ref{14}) can be applied to any 
operators. For instance, to the density operators, or reduced density matrices, and 
to correlation operators, whose matrix elements are correlation functions. This 
generalization of the entanglement can be called {\it correlation entanglement}.  
The correlation entanglement should not be confused with the so-called generalized
entanglement \cite{Barnum_19,Viola_20,Barnum_21} that considers a kind of entanglement
based of algebraic properties of states.  

Generally, quantum statistical systems can be fully characterized by reduced density 
matrices \cite{Coleman_22}. In particular, they contain all information on quantum 
correlations in the system. Let us start with the first-order density matrix
\be
\label{43}
 \rho_1(x,x') \; = \; {\rm Tr} \; \psi(x) \; \hat\rho \; \psi^\dgr(x') \; = \;
\lgl \; \psi^\dgr(x') \; \psi(x) \; \rgl \;  ,
\ee
where the angle brackets imply the statistical averaging in the Fock space, with a
statistical operator $\hat{\rho}$, and with $x$ representing the set of the appropriate 
variables, such as spatial coordinates, spin etc. We may consider the above expression 
as a matrix element, with respect to the variable $x$, of the first-order density
operator
\be
\label{44}
\hat\rho_1 \; = \; [\; \rho_1(x,x') \; ] \;   .
\ee

This operator acts on the Hilbert space
\be
\label{45}
 \cH_1 \; = \; \overline\cL \{\; |\; k \; \rgl \; \}  
\ee
that is a closed linear envelope over a basis formed by the functions
\be
\label{46}
|\; k \; \rgl \; \equiv \; |\; \vp_k \; \rgl \; = \; [\; \vp_k(x) \; ]
\ee
considered as columns with respect to $x$, with $k$ being a quantum multi-index. 
As usual, the basis is assumed to be orthonormalized,
$$
\lgl \; k \; | \; p \; \rgl \; = \; \int \vp_k^*(x) \; \vp_p(x) \; dx \; = \;
\dlt_{kp} \;  .
$$
The matrix elements of $\hat{\rho}_1$ with respect to the multi-indices are
$$
\lgl \; k \; | \; \hat\rho_1\; | \; p \; \rgl \; = \; 
\int  \vp_k^*(x) \; \rho_1(x,x') \; \vp_p(x') \; dx dx' \;   .
$$

The Hermitian norm of the density operator (\ref{44}) reads as
\be
\label{47}
 ||\; \hat\rho_1\; || \; = \; 
\sup_k  \lgl \; k \; | \; \hat\rho_1\; | \; k \; \rgl \;  .
\ee
The trace over the Hilbert space (\ref{45}) is
\be
\label{48}
{\rm Tr}_{\cH_1} \; \hat\rho_1 \; = \; \sum_k  
 \lgl \; k \; | \; \hat\rho_1\; | \; k \; \rgl \; = \;
\int \rho_1(x,x) \; dx \; = \; N \; .
\ee

One can define the density matrix of second order
\be
\label{49}
\rho_2(x_1,x_2,x_1',x_2') \; = \; 
\lgl \; \psi^\dgr(x_2') \; \psi^\dgr(x_1') \; \psi(x_1) \; \psi(x_2) \; \rgl
\ee
enjoying the symmetry property
\be
\label{50}
\rho_2(x_1,x_2,x_1',x_2') \; = \; \rho_2(x_2,x_1,x_2',x_1') \;   .
\ee
Treating this as a matrix element in variables $x$, one has the density operator
\be
\label{51}
 \hat\rho_2 \; = \; [\; \rho_2(x_1,x_2,x_1',x_2') \; ] \;  .
\ee
         
The operator (\ref{51}) acts on the Hilbert space
\be
\label{52}
\cH \; = \; \cH_1 \bigotimes \cH_1 \; = \; 
\overline\cL\{ \; |\; k p \; \; \rgl \; \}
\ee
that is a closed linear envelope over the basis functions
\be
\label{53}
|\; k p \; \; \rgl \; = \; 
|\; \vp_k  \; \; \rgl \; \bigotimes \;  |\; \vp_p  \; \; \rgl \; = \; 
[\; \vp_k(x) \;] \; \bigotimes \; [\; \vp_p(x') \;] \;  ,
\ee
being the natural orbitals for the considered system \cite{Coleman_22}. The matrix 
elements of the latter operator with respect to the multi-indices are
$$
\lgl \; k p \; | \; \hat\rho_2 \; |\; k' p' \; \; \rgl \; = \; 
\int \vp_k^*(x_1) \; \vp_p^*(x_2) \;
\rho_2(x_1,x_2,x_1',x_2') \; \vp_{k'}(x_1') \; \vp_{p'}(x_2') \; 
dx_1 dx_2 dx_1' dx_2 ' \;  .
$$  

The norm of the density operator (\ref{51}) is
\be
\label{54}
 || \; \hat\rho_2 \; || \; = \; 
\sup_{kp} \; \lgl kp \; | \; \hat\rho_2 \; | \; kp \; \rgl  
\ee
and the trace over space (\ref{52}) is
\be
\label{55}
{\rm Tr}_\cH \; \hat\rho_2 \; = \; 
\sum_{kp} \; \lgl kp \; | \; \hat\rho_2 \; | \; kp \; \rgl \; = \; 
\int \rho_2(x_1,x_2,x_1,x_2) \; dx_1 dx_2 \;  .
\ee

Partial traces are defined as 
\be
\label{56}
 {\rm Tr}_{\cH_1} \hat\rho_2 \; = \; 
\sum_{k} \; \lgl k \; | \; \hat\rho_2 \; | \; k \; \rgl \; = \; \hat R \; = \; 
[\; R(x,x') \; ] \;  ,
\ee
with the reduced operator $\hat{R}$ possessing the matrix elements
\be
\label{57}
  R(x,x') \; = \;  \int \rho_2(x_1,x,x_1,x') \; dx_1 \; = \; 
\int \rho_2(x,x_2,x',x_2) \; dx_2 \; .   
\ee
The norm of this operator reads as 
\be
\label{58}
|| \; \hat R \; || \; = \; 
\sup_k \; \lgl k \; | \; \hat R \; | \; k \; \rgl \; = \; 
\sup_k \int \vp_k^*(x) \; R(x,x') \; \vp_k(x') \; dx dx'
\ee
and the trace is
\be
\label{59}
{\rm Tr}_{\cH_1} \hat R \; =  \; {\rm Tr}_\cH \; \hat\rho_2 \; .
\ee

The factorized nonentangling operator (\ref{11}) takes the form
\be
\label{60}
 \hat\rho_2^\otimes \; = \; 
\frac{\hat R \bigotimes \hat R}{{\rm Tr}_\cH \;\hat\rho_2} 
\qquad  
({\rm Tr}_{\cH}\; \hat\rho_2^\otimes = {\rm Tr}_{\cH}\; \hat\rho_2) \; .
\ee
Hence, its norm becomes
\be
\label{61}
||\; \hat\rho_2^\otimes \; || \; = \; 
\frac{||\; \hat R \; ||^2}{{\rm Tr}_\cH \; \hat\rho_2} \;   .
\ee

Finally, the measure of entanglement production (correlation entanglement) is
\be
\label{62}
 \ep(\hat\rho_2) \; = \; 
\log \; \frac{||\; \hat\rho_2 \; ||}{||\; \hat\rho_2^\otimes \; ||} \;  .
\ee

In a similar way, it is possible to introduce the measure of quantum correlations for 
any other correlation operator
\be
\label{63}
 \hat C_2 \; = \; [ \; C_2(x_1,x_2,x_1',x_2') \; ] \;  ,
\ee
whose matrix elements are the correlation functions 
\be
\label{64}
  C_2(x_1,x_2,x_1',x_2') \; = \; 
\lgl \; \hat A^+(x_2') \; \hat A^+(x_1') \; \hat A(x_1) \; \hat A(x_2) \; \rgl \; ,
\ee
formed by any operators of local observables. The measure
\be
\label{65}
\ep(\hat C_2) \; = \; \log \; \frac{||\; \hat C_2 \; ||}{||\; \hat C_2^\otimes \; ||}
\ee
quantifies the amount of correlation entanglement in the correlation operator.

\section{Conclusion}

The notion of entanglement production is suggested. The measure quantifying this 
notion is described. The measure can be used for characterizing any operators and
any systems, simple and complicated, bipartite and multipartite, equilibrium
and nonequilibrium. The application of the measure is illustrated by the examples 
of quantum-mechanical systems with pure and mixed states. The generalization of the
entanglement production by statistical operators to the correlation entanglement
of reduced density operators and correlation operators is presented. As other examples 
of application, we can mention multimode coherent states of Bose condensate in optical 
lattices \cite{Yukalov_23,Yukalov_24,Yukalov_25,Yukalov_26}, spin correlation operators
\cite{Yukalov_11,Yukalov_12,Yukalov_16}, and pseudospin operators employed in the theory
of coherent radiation in nonequilibrium systems \cite{Yukalov_16,Yukalov_27}. The change
of the correlation entanglement under phase transitions has been studied in Refs.
\cite{Yukalov_12,Yukalov_16}. Entanglement is an important resource for quantum computing
and quantum information processing \cite{Williams_1,Nielsen_2,Vedral_3,Keyl_4,Horodecki_5,
Guhne_6,Yukalov_7,Wilde_8,Bernhardt_28,Hidari_29}. This explains the interest to 
entanglement production under quantum measurements in quantum decision theory 
\cite{Yukalov_30,Yukalov_31}.

\vskip 2cm

{\parindent=0pt
{\bf Funding}

\vskip 2mm
This work was supported by ongoing institutional funding. No additional grants to carry 
out or direct this particular research were obtained.

\vskip 5mm

{\bf Conflicts of interest}

\vskip 2mm
The authors of this work declare that they have no conflicts of interest.

}

\end{document}